\documentclass[aps,prappl,twocolumn,superscriptaddress]{revtex4}
\usepackage[intlimits]{amsmath}
\usepackage{amsfonts}
\usepackage{graphics}
\usepackage{subfigure}
\usepackage[usenames]{color}
\usepackage{epstopdf,epsfig}
\usepackage{color}
\usepackage{indentfirst}
\usepackage{graphicx}% Include figure files
\usepackage[utf8]{inputenc}
\newcommand{\ket}[1]{\vert #1 \rangle}

\newcommand{\ave}[1]{\langle #1 \rangle}
\newcommand{\ketbra}[2]{|#1\backslash\!\!\!/#2|}

\newcommand{\Tr}[1]{\text{Tr}\left(  #1 \right)}
\newcommand{\Eq}[1]{$\text{Eq.}\left(  #1 \right)$}
\usepackage{bbold} %for the identity matrix
\usepackage[usenames]{color}
  \expandafter\let\csname equation*\endcsname\relax
  \expandafter\let\csname endequation*\endcsname\relax
 \usepackage{dsfont}
\usepackage{soul}
\begin{document}

\title{Thermal transistor effect in quantum systems}

\author{ Antonio Mandarino}
\affiliation{Institut UTINAM - UMR 6213, CNRS, Universit\'{e} Bourgogne Franche-Comt\'{e}, Observatoire des Sciences de l'Univers THETA, 41 bis avenue de l'Observatoire, F-25010 Besan\c{c}on, France}
\author{ Karl Joulain}
\affiliation{Insitut Pprime, CNRS, Universit\'{e} de Poitiers, ISAE-ENSMA, F-86962 Futuroscope Chasseneuil, France}
\author{ Melisa Dom\'{i}nguez G\'{o}mez}
\affiliation{Insitut Pprime, CNRS, Universit\'{e} de Poitiers, ISAE-ENSMA, F-86962 Futuroscope Chasseneuil, France}
\affiliation{Instituto de Fisica, Universidad de Antioquia, Calle 70 No. 52-21, Medellin, Colombia}
\author{ Bruno Bellomo}
\affiliation{Institut UTINAM - UMR 6213, CNRS, Universit\'{e} Bourgogne Franche-Comt\'{e}, Observatoire des Sciences de l'Univers THETA, 41 bis avenue de l'Observatoire, F-25010 Besan\c{c}on, France}

\begin{abstract}
\noindent

We study a quantum system composed of three interacting
qubits, each coupled to a different thermal reservoir. We show how to engineer it in order to
build a quantum device that is analogous to an electronic bipolar transistor.
We outline how the interaction among the qubits plays a crucial role for the
appearance of the effect, also linking it to the characteristics of system-bath
interactions that govern the decoherence and dissipation mechanism of the system.
By comparing with previous proposals, the model considered here extends the
regime of parameters where the transistor effect shows up and its robustness
with respect to small variations of the coupling parameters.
Moreover, our model appears to be more realistic and directly connected in
terms of potential implementations to feasible setups in the domain of quantum spin chains  and molecular nanomagnets.

\end{abstract}
\maketitle
\section{Introduction}
The exploitation of  classical thermodynamics
has conducted to the technological revolution that shaped our world		
since the Nineteenth century \cite{Carnot}. In recent years the
exploration of the thermodynamics of quantum systems has
given birth to quantum thermodynamics: a high impact research field
either on the fundamental level and the applicative one \cite{QthermoBook}.
In analogy to classical models,
quantum  heat engines have been proposed and realized \cite{eng1, eng2, eng3}.
In particular, the properties of non-equilibrium open quantum systems has
been employed to study how to obtain many-body entanglement \cite{BA15} or
efficient  flux management \cite{LBA15}.
Moreover, the implementation of  non-equilibrium quantum heat machines has been discussed in
solid state set-ups \cite{CPF15} or in more  complex scenarios \cite{Ali15}.

The ability to manipulate quantum resources is a demanding
but fruitful task that has to be pursued in order to build
novel devices. A potential high impact apparatus will be the one aiming at
the control of thermal energy transport in quantum systems,
and at the  amplification of heat fluxes among the different parts
constituting a composite system.
A promising approach in the aforementioned task is to pursue
the flourishing path followed in electronics after the
realization of rectifiers and transistors with semi-conducting materials  \cite{BB48}
that paved the way to build logic gates and to the information and computational technology \cite{millman}.

Typically, thermal and electric currents phenomena are  empirically well described
by the Fourier's and Ohm's laws, respectively. In fact, the functional dependence
upon the two respective control variables, temperature and voltage, is of the same type.
Recently, evidences of the emergence of the Fourier's law
has been pointed out also in the quantum realm \cite{Qfourier1, Qfourier2}.

The proposal of such thermal analogues of the electronic  rectifier and transistor
is based on the assumption that in a thermal set-up
the role played by batteries could be played
by thermostats. This analogy allowed to come forward with
various proposals of mesoscopic thermal devices such as
rectifiers and diodes \cite{TPC02, LWC04} and
transistors \cite{LLC06}, the latter operating
either in near field \cite{NearField} or in far
field \cite{FarField}  regime of thermal radiation,
and also thermal logic gates \cite{gate, RevPhononics}.

The theoretical efforts are bolstered by the experimental ingenuity
involving a wide range of experimental platforms ranging from
carbon  and boron nitride nanotube structures \cite{carbnano} and bulk   oxide materials \cite{oxmat}
to semiconductors quantum dots \cite{qdots},  magnonic systems \cite{DiodeMag}
and phase changing materials, such as vanadium dioxide VO$_2$ \cite{vanad1, vanad2}, and   ceramics   materials that below a critical temperature behave
like high temperature superconductors and  above it as dielectrics  \cite{ceram1, ceram2}.

In the same time, the quest of atomic-scale devices
for quantum computational purposes  is increasingly pushing
the necessity to control and manipulate single or few atoms systems.
In particular, the isolation and coherent manipulation of single spins,
that are one of the best candidates for a non-optical implementation of a single qubit,
is achieved using optical traps and electrical techniques \cite{manip}.
Recently, several novel quantum
technology devices have been proposed, including
isolators based on photonic transitions \cite{qtech1},
rectifiers \cite{qtech2, qtech3}, transistors
in  an electromagnetic controlled environment \cite{qtech4, qtech5},
and also phonon-thermoelectric transistors \cite{qtech6}.

Moreover, quantum systems suffer of an unavoidable coupling
to  their environment,   typically modeled as a thermal reservoir of much
bigger dimensions of those of the system \cite{BP02, Schaller}.
Despite some detrimental phenomena related to decoherence and dissipation,
control  and engineering of interacting quantum systems coupled
with different thermal baths have led to several studies pointing out
how to build  the smallest thermal refrigerator \cite{refr1, refr2},
how to rectify a thermal current at the very quantum level
\cite{QTD1,QTD2, QTD3} and recently the building blocks of a
quantum thermal transistor have been discussed considering
an integrable spin-chain model \cite{QTT} and in a qubit-qutrit system \cite{QTT2}  .

In this paper we analyze the ability to design a quantum thermal
transistor employing a  spin ring that can, in principle,
be implemented on a molecular nanomagnet \cite{NanoMag}. In comparison to a previous model \cite{QTT},
we consider a more complex scenario where the collective quantum behavior of few spins is expected to play a  crucial role in the functioning of the device.

The paper is structured as follows. In Section~\ref{Model}, we examine the  spin-ring model
outlining the frameworks where the theoretical model
finds a direct implementability. This section is also 
devoted to the quantum dissipative process that
allows the system to exchange heat fluxes.
In Section~\ref{QTT}, we discuss how the proposed systems
can be adopted to build a quantum thermal transistor
and some limits of their applicability. 
Having in mind a direct experimental implementation, 
in Section~\ref{Robust} the robustness  of the transistor effect
against unavoidable fluctuations and perturbations is addressed.
Section~\ref{Concl} concludes the paper with some final remarks and prospects.
Some details can be found in the Appendix.

 \section{The spin-ring model}
\label{Model}
Quantum critical  spin chains  have longly been considered
as  one of the best substrates to test and implement
the future quantum computing and the novel quantum devices.
Their main characteristic is to show wide versatility \cite{RevQC}.
In fact, considering the most general model for quantum magnetism,
described by the Heisenberg Hamiltonian,
one can explore a great variety of universality class
by an appropriate tuning of the interaction strengths.
Moreover, recent modern developments allow to realize  spin chains of few
atoms having net spin $s=1/2$,
opening the road to the exploitation of the physics
of some well known models, viz., Ising, XY   and XXZ  ones
\cite{chain1, chain2, ExpIsing1,ExpIsing2,ExpIsing3}.

In parallel, molecular nanomagnets \cite{NanoMag}
are also attracting more interest as a feasible platform
for quantum technological purposes \cite{TZ13}.
Specifically, molecular nanomagnets,  composed by $N=3$
main units, have been proved to be useful for quantum
thermometry \cite{LMG14} and for coherent-manipulation of
three-qubit states \cite{3QMag}.

\begin{figure}[t]
\includegraphics[width= 0.80 \columnwidth]{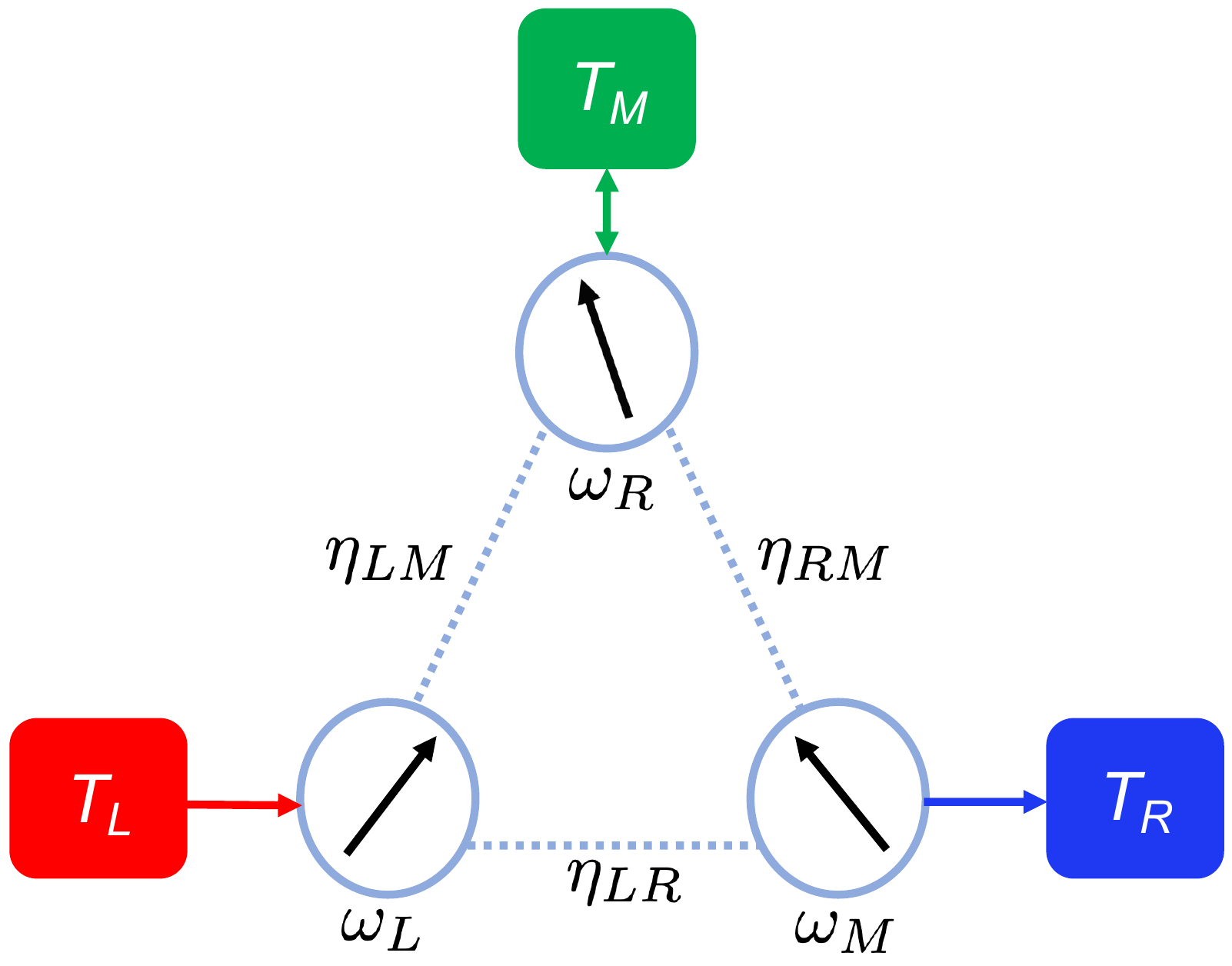}
\caption{\label{fig:1}
(Color online) Schematic picture of a
system composed by three qubits interacting each one with any
other according to the model in \Eq{\ref{eq:hsys}}.
Each of them is dissipating into a different
thermal environment as described in  \Eq{\ref{eq:hamiltonians}}.   }
\end{figure}

Spin chains and nanomagnets are theoretically well described
by the same effective Hamiltonian upon fixing the configuration geometry.
Here, we consider a system of three two-level systems (qubits)
of frequency $\omega_p $  $(p=L,M,R)$, embedded in a magnetic field,
 and we introduce the vector of the effective spin operators of any qubit
$\hat{\mathbf{S}}_{p} = (\sigma^x_p,\sigma^y_p,\sigma^z_p)^T $
and  the 3-by-3 matrix $ \boldsymbol{\mathsf{\lambda}} $
governing the coupling  between spins along different polarization axis,
where $\sigma^{i}_p$  ($i=x,y,z$) is the $i$-th Pauli matrix of the $p$-th qubit.
A realistic description of a triangular spin chain
with a qubit in each vertex labeled  $L$, $M$ and $R$,  as depicted in Fig.~\ref{fig:1},
is given by the following Hamiltonian  (hereafter  the  reduced Plank constant and the Boltzmann constant are set equal to one, $\hbar=k_{\textup{B}}=1$):
\begin{equation}
H_{S}= \frac12 \sum_{p=L,M,R} \omega_p \sigma^z_p + \sum_{p \neq q} \hat{\mathbf{S}}_p^T \boldsymbol{\mathsf{\lambda}}\hat{\mathbf{S}}_q.
\end{equation}

The first term is the sum of single particle \textit{free}
Hamiltonians of three qubits immersed in a magnetic field
pointing along the direction of the $z-$axis and it  is
responsible of the non-zero field splitting
of each qubit, while the second term is the intra-chain spin-spin exchange coupling.
Such an Hamiltonian takes into account symmetric and antisymmetric exchange
interactions. A proper tuning of $ \boldsymbol{\mathsf{\lambda}} $
can give a situation where only the former type of interactions occurs,
like in the standard Heisenberg Hamiltonian, or can introduce a coupling
between non-homologous components of the spin operators,
as in the Dzyaloshinskii-Moriya interaction \cite{Dz58,Mor60}.
We omit here the overcomplexity
introduced by an antisymmetric exchange, hence we assume that
the tensor $\boldsymbol{\mathsf{\lambda}}$ contains only symmetric terms.

This allows us to consider spins interacting among them
through an Heisenberg type Hamiltonian. Therefore, specifying
their couplings $\eta_{pq}$ with $p,q= L, M,R$  and $p\neq q$,
and the strengths along different polarization axis
$\lambda^{i}$, the Hamiltonian of the system reads:
\begin{equation}
\label{eq:hsys}
H_S = \frac12 \sum_{p=L,M,R}  \omega_p \sigma^z_p + \frac12 \sum_{p \neq q} \eta_{pq} \sum_{i=x,y,z} \lambda^{i} \sigma^i_p \sigma^i_q,
\end{equation}
where  we have chosen equal $\lambda^{i}$ for the three qubits.

This Hamiltonian describes the simplest
non-trivial example of spin ring whose properties
can be observed with good approximation in a $\{\text{Cu}_3\}$-Type
nanomagnet, a complex having an almost equilateral triangular shape
with a $\text{Cu}^{2+}$ ion of spin$-1/2$ at each vertex \cite{Cu3}.

\subsection{Comparison between two Ising-type rings}
\label{Comparison}
Joulain et al. in their seminal paper \cite{QTT}
have addressed how a thermal transistor  can be realized with a quantum
system of three interacting qubits, coupled to a thermal reservoir each,
and showed how it is analogous to an electronic bipolar one.
Their discussion is based on a peculiar choice of a very simple model,
that implies a lot of strong assumptions on the underlying physics.

Let us refer to the model adopted in \cite{QTT} as the $z-$axis.
In fact, there the magnetic field applied on the three qubits
is longitudinal to the direction of the spin-spin interactions.
Being more specific, the exchange interaction in $H_S$  of  \Eq{\ref{eq:hsys}}
is non-zero only along the same direction of the magnetic field,
i.e. only $\lambda^{z} \neq 0$.
Among the other assumptions, this is the one
that, in our opinion, requires now a detailed discussion.
In fact, this premise, along with some other minor approximations,
has the advantage to let the model and the dissipative dynamics
be solvable with analytical techniques, since  the operators
$\sigma_p^z\sigma_q^z$  commute with the free Hamiltonian.
Essentially, the only effect of the zero field splitting terms $\sigma_p^z\sigma_q^z$ 
is  to modify the eigenenergies of the system,
but the collective eigenfunctions  remain the same and can be still expressed as
a tensor product of single-particle eigenstates.
We note that this kind of spin ring, even if  considered in a quantum context,
is a semi-classical model. In fact, the spin$-1/2$ operators $\sigma^z_p$
can be alternatively viewed as a classical magnetic moment
pointing only along two opposite directions.
The $z-$axis Ising model, hence, takes correctly into
account a quantization of the magnetic moment, but it does not
consider quantum fluctuations due to other spatial components
of the same spin operator $\hat{\mathbf{S}}_p$
and any possible critical behavior is disregarded.

It is worth noting that in order to propose a truly quantum model
a more complex interaction should be taken into account.
The reason why we discuss the class of Hamiltonians
defined in  \Eq{\ref{eq:hsys}}, where, in general, the sum of the  single particle  contributions
does not commute with the interaction term is twofold:
(i) from a purely theoretical point of view,  these systems may exhibit
strongly correlated and nonlocal eigenstates, opening the
door to phenomena with an enhancement due to quantum correlations or criticality \cite{RevQC};
(ii) from an experimentally inspired perspective, nowadays,
the control of nanomagnets,  spin chains or even
quantum dot molecules, has been well established
for quantum critical systems \cite{ExpIsing1, ExpIsing2, ExpIsing3}.

For the aforementioned reasons our aim is  then to propose
a well  established quantum setting, richer than the one considered in \cite{QTT}, where the thermal transistor effect displays
and  to study its robustness with respect to the operating temperature
and against spurious fluctuations of the exchange interactions.

In particular,   we will mainly focus on the case of magnetic systems
that  belong to the universality class of the
quantum critical Ising model in a transverse field and
we will refer shorthand to it as the $x-$axis Ising model.
It can be realized considering in  \Eq{\ref{eq:hsys}} that
the coupling among the qubits acts only along a direction  (that we name $x$)
transverse to the applied magnetic field,  i.e. only $\lambda^{x} \neq 0$.
We remark that once the coupling with the environment is added (see next sub-section),
the different directions transverse to the applied magnetic field
are not equivalent anymore.
We  address  in  Section~\ref{Robust}
the presence of perturbations on both the $x$ and $z$ configurations.

The $x-$axis Ising model introduced as the exemplary
model to study quantum phase transitions in magnetic
systems and  their critical behaviors has been employed
in a wide range of protocols to characterize and
fulfill quantum computational tasks \cite{RevQC}.
The main feature of such a model resides in the fact
that its collective states are no longer a tensor product
of the single particles states, but are non-separable.
Its pivotal role for the comprehension of magnetism at the quantum scale
has fostered the research of a model magnet where its properties can
be experimentally measured with different techniques.
Nowadays, the most germane realization has been achieved in
the low-energy magnetic excitation of the insulator LiHoF$_4$ \cite{ExpIsing1}
or with crystals of the ferromagnetic CoNb$_2$O$_6$
where the  spin resides on the Co$^{2+}$ ion \cite{ExpIsing2}.
A detailed presentation of the spectrum of the $x-$axis Ising
model for the purposes of this paper is shown in Appendix \ref{ApA},
while a complete discussion of the spectrum of $z-$axis one  can be found in \cite{QTT}.

\subsection{Dissipative dynamics}

Each qubit composing the spin ring described above
 suffers of an unavoidable coupling with its surrounding environment,
that  is modeled as a thermal bosonic reservoir.
The temperatures of the three reservoirs are, in general, different,
giving rise to an out of thermal equilibrium scenario where the temperatures are meant to be tunable at will.
The total Hamiltonian is then $H =  H_S+H_B+H_I$, where
the bath and the system-bath Hamiltonians, respectively, read:
\begin{equation}
\label{eq:hamiltonians}
\begin{split}
H_B &= \sum_{p=L,M,R} \sum_{k} \omega_k a_k^{p\,\dagger}a_k^{p} \\
H_I &= \sum_{p=L,M,R} \sigma_p^x \otimes \sum_k  g_k^p  \left(a_k^{p\,\dagger}+a_k^{p}\right),
\end{split}
\end{equation}
 where $a_k^{p}$  and $a_k^{p\,\dagger}$ are the bosonic
 operators of bath $p$ and $g_k^p$  are the coupling strengths.
A schematic representation of the  global  system is sketched in Fig.~\ref{fig:1}.
As spin ring Hamiltonian $H_S$, we consider the one of  \Eq{\ref{eq:hsys}}.

The dynamics of the 3-qubit system is dissipative and,
under the Born-Markov  and secular approximations \cite{BP02},
the evolution of the density matrix is described
by a master equation of the following form:
\begin{equation} \label{eq:master}
\dot{\rho}=-i[H_S+H_{LS},\rho] +\sum_{p=1}^3 \mathcal{L}_p[\rho],
\end{equation}
where $H_{LS}=\sum_{p,\,\omega} s_p(\omega) A_p^{\dagger}(\omega) A_p(\omega) $
is the Lamb shift Hamiltonian,
$A_p(\omega)=\sum_{\omega=\epsilon_i-\epsilon_j} \ketbra{\epsilon_j}{\epsilon_j}\sigma_p^x\ketbra{\epsilon_i}{\epsilon_i}$,
 $ \ket{\epsilon_i}$ with $i=  1, ..., 8 $
are the dressed states of the Hamiltonian $H_S$, as described
in the Appendix \ref{ApA}, and the Lindblad operators are given by \cite{BP02}

\begin{equation}
\label{eq:lindblad}
\mathcal{L}_p[\rho]=\sum_{\omega} \gamma(\omega) \Bigl[ A_p(\omega)\rho A_p^\dagger(\omega) -\frac12  \{\rho, A_p^{\dagger}(\omega) A_p(\omega)\} \Bigr]
,\end{equation}
where  $\gamma_{p}(\omega)=\mathcal{J}_p(\omega)[1 +n_p(\omega)] $ for $\omega>0$
and $\gamma_{p}(\omega)~=~\mathcal{J}_p(|\omega|) n_p(|\omega|) $ for $\omega<0$.
The average number of excitations
and the spectral density of the $p$-th reservoir  are respectively
$n_p(\omega)=[\mathrm{exp} \frac{\omega}{T_p}-1]^{-1}$ and  $\mathcal{J}_p(\omega)$.
In all the numerical computations, we will choose Ohmic reservoirs
characterized by a linear spectral density $\mathcal{J}_p(\omega)= \kappa_p \omega$ (at least in the range of working frequencies),
using  values of $\kappa_p$ different for the three baths in order to suitably
engineer the effect of the environments on the three qubits. 
It follows that $\gamma_p(0)=\mathrm{lim}_{\omega\rightarrow 0^+}\mathcal{J}_p(\omega) n_p(\omega) = \kappa_p
T_p$. In the Appendix \ref{ApA}, a scheme of the transitions induced by the three thermal baths is also reported.

We remark that as we will consider a strong coupling between the three qubits
(comparable or higher than the bare frequencies $\omega_p$),
we  need the above microscopic approach  to derive the Lindblad operators
responsible of the non-unitary evolution of the density matrix,
avoiding any extra simplification that can be somehow justified
when the spin-spin interaction is  much smaller than
the bare frequencies of the qubits \cite{BP02, QTD1}.
In this limit, we cannot consider phenomenological master equations where the operators
$A(\omega)$ are built starting from the eigenstates of the bare system  Hamiltonian.
For the $z-$axis Ising model, these coincide with the dressed eigenstates,
while for the $x-$type this is not the case.

We also note that since $H_{LS}$ commutes with $H_S$, it only leads to a renormalization of
the unperturbed energy levels of $H_S$ induced by the coupling
with the reservoirs and that the steady state is independent on it.

The three temperatures of the environment will be in general chosen
to be different, leading to out of thermal equilibrium steady states,
which are not expected  to be three-qubit thermal states.
At the same time, we expect that each qubit will not thermalize
(its own reduced state) to the temperature of its own reservoir
because of the collective nature of the quantum dynamics.

\section{Quantum thermal transistor}
\label{QTT}
The particular operating principle of an electronic transistor makes possible to regulate  the
currents at two of its terminals, that could take also very high values,
regulating a much smaller current   injected through a third terminal.
This peculiarity made the transistor particularly suitable to build logic gates
commonly used in nowadays electronic.
We propose here a thermal analogue of this device able to amplify
some of the heat currents circulating in a system as in Fig.~\ref{fig:1}.
The control parameter that will play the role of the gate potential
in an electronic transistor is the temperature $T_M$ of the reservoir coupled
to the \emph{middle} qubit of frequency $\omega_M$.

\subsection{Heat currents in a thermal transistor}
\label{QTT:A}
A straightforward way to take into account  the heat flow in and out a quantum system \cite{kosloff},
is to link the variation of its mean energy $\ave{H_s}$
to the sum of the heat currents $J_p$ exchanged between the quantum system
and each bath (no power comes from other external
sources in our model):
% \cite{fourier}:
\begin{equation}
\sum_p J_p =  \frac{ \partial\ave{ H_S}}{\partial t}.
\end{equation}
%\frac{ \partial\ave{ H_S}}{\partial t} = \Tr{\dot{\rho} H_S} =
By substituting \Eq{\ref{eq:master}} in the previous expression, one can compute
the current $J_p$ that each thermostat exchanges with the system as
\begin{equation}
\label{currJ}
J_p = \Tr{\mathcal{L}_p[\rho] H_S}.
\end{equation}
We note  that at the steady state $\rho_{ss}$,
it holds  $\dot{\rho}_{ss} =0$, so that  the total energy is conserved in time. This implies that $\sum_p J_p = 0$.
A minor remark concerns the fact that  the Lamb shift term appearing in
\Eq{\ref{eq:master}} does not contribute neither to the final expression of $J_p$
neither to the steady state, so that the steady currents are independent on it.
In the following, we always address steady configurations and we use $J_p$ to indicate steady currents.

Following the geometry of the configuration of Fig.~\ref{fig:1},
we will refer to \emph{left}, \emph{middle} and \emph{right}  currents ($J_L$, $J_M$ and $J_R$) to those
exchanged, collectively, between the  spin ring and, respectively, the left, the middle and the right thermostat.
From now on, we assume that the hot reservoir, providing
the energy to the system, is the left one, whereas the
cold one, that adsorbs the heat flux, is the one placed to the right,
and finally the bath in the middle acts as a control.
Borrowing the familiar terminology of the bipolar transistor,
the right qubit is playing the role of the emitter, while the left
one of the collector and the middle one is indeed the analogue of the base.

The thermal transistor effect happens when a small change in the
the control temperature $T_M$ produce a significant variation of
the two lateral currents in contrast with a tiny variation of $J_M$.
To study the occurrence of this effect
we define the differential thermal resistances
\begin{equation}
\label{ThermalRes}
\chi_{s} = \left(\frac{\partial J_{s}}{\partial T_M} \right)^{-1}_{T_{s} = \text{const}}, \quad s=L, R,
\end{equation}
and, as in the spirit of the electronic transistor \cite{BB48}, we introduce
a dynamical amplification factor $\alpha_s$,
function of the control temperature $T_M$, defined as
\begin{equation}
\label{alpha}
\alpha_{s} = \frac{ \partial J_{s}}{\partial J_M}=-\frac{\chi_{s}}{\chi_{L}+\chi_{R}}.
\end{equation}

The adimensional parameters in \Eq{\ref{alpha}} are the figures of merit used
to have a quantitative benchmark of the presence of the thermal transistor effect
and they satisfy the relation $ \alpha_{L} + \alpha_{R} = -1$.
In particular, for regions where $|\alpha_s|\gg 1$
one can infer that we are in presence of a strong amplification of the
lateral currents in comparison to that controlled by the heat bath in central position.
Therefore, it is evident that to amplify the thermal currents
one of the two differential thermal resistances should be negative.
We underline that this condition was already discussed in the
context of non-linear lattices \cite{LLC06},
three-terminal graphene devices \cite{graphene} and  band structured engineered
silicene superlattices \cite{silicene}.

\subsection{Thermal transistor in an $x-$axis Ising model}\label{sec:IIIB}
As said in Sec.~\ref{Comparison} the main case we want to discuss is when the three
qubits are described by an $x-$axis Ising model.
In particular, we choose open boundary condition, i.e. $\eta_{LR}=0$.

\begin{figure}[t]
\includegraphics[width= 0.65 \columnwidth]{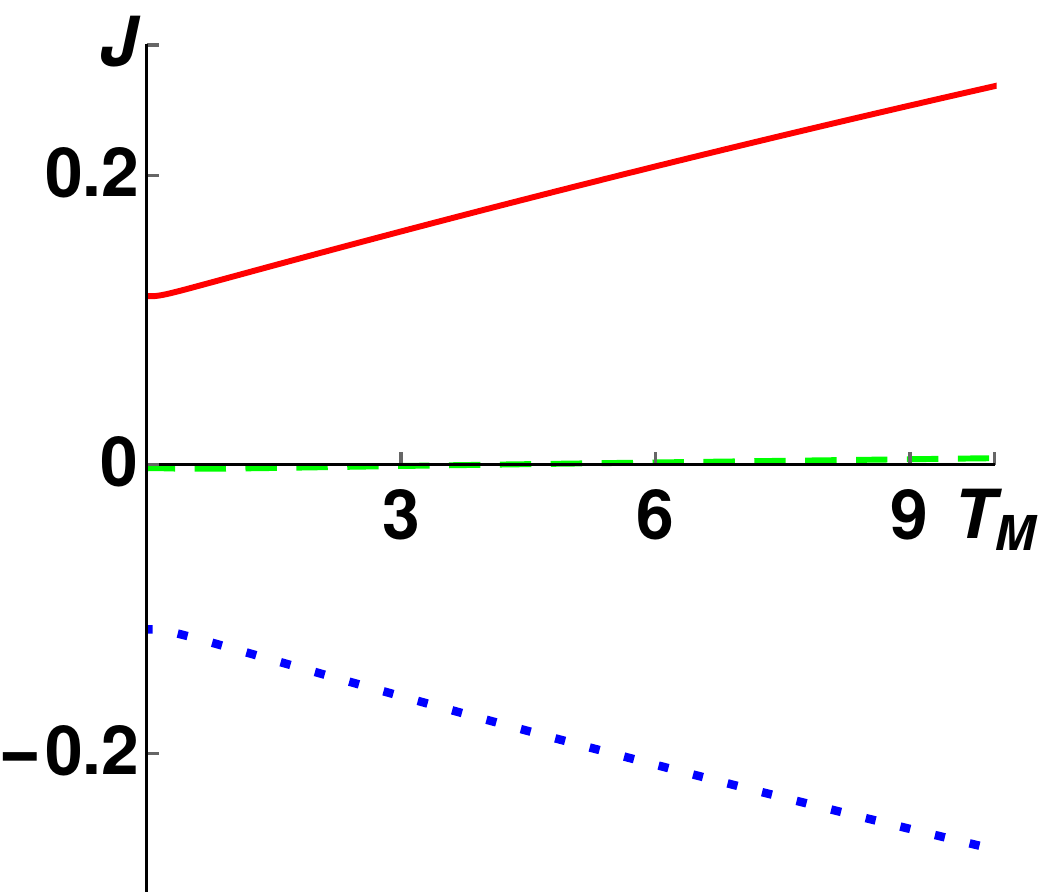}
\caption{\label{fig:2}
(Color online) The three thermal currents defined in \Eq{\ref{currJ}} (in units of $\kappa \  \Delta^2$) exchanged by the
environments with the system  as a function of the control parameter $T_M$ (in units of  $\Delta$),
when the system is described by an $x-$axis Ising model.
The solid red line refers to $J_L$, the dashed green line to $J_M$
and the blue dotted one to $J_R$.
The parameters are $\omega_L= \Delta, \omega_M= 0.1 \ \Delta,   \omega_R= 0.2 \ \Delta$,
$\eta_{LM}=\eta_{MR}=\Delta, \, \eta_{LR}=0$,
$\kappa_L=\kappa_M = \kappa,  \kappa_R= 100 \ \kappa $,
$T_L= 10 \ \Delta$ and $T_R = 0.01\ \Delta $. 
In the following figures, currents and temperatures are always meant, respectively, in units of $\kappa \  \Delta^2$ and $\Delta$.
} 
\end{figure}

In Fig.~\ref{fig:2} we show the behavior of each current.
In order to deal with adimensional quantities we plot the currents $J_p$ in units of 
$\kappa \Delta^2$, where $\kappa=\kappa_L$ is the parameter characterizing 
the strength of the dissipation of the left qubit and $\Delta = \omega_L$,  $\omega_L$ 
being  assumed as the reference frequency. The currents are function of the temperature $T_M$, 
given in units of $\Delta$. We observe that once expressed also $\kappa_M$ and $\kappa_R$ in units of $\kappa$,
the steady state of \Eq{\ref{eq:master}} is invariant with respect to $\kappa$, while the currents are just proportional to it.
Its value must just be small enough to guarantee the validity of
the approximations used to derive the master equation of  \Eq{\ref{eq:master}}. 
In the following, all the values of temperatures are given in units of $\Delta$, while for currents they are in units of $\kappa \  \Delta^2$.
It is possible to appreciate  in Fig.~\ref{fig:2} how the thermal transistor effect manifests  itself
and how the amplification of $J_L$ and $J_R$ is continuously achieved.

We stress here the  crucial role  played by the non-equilibrium steady state.
In fact, the presence of three environments at different temperatures
plugged to distant sites of a magnetic system  is the responsible
of the three heat flows circulating into it.
In particular, it is not surprising observing
that even at $T_M =0$ each of the currents has a non zero value.

\begin{figure}[t]
\includegraphics[width= 0.65 \columnwidth]{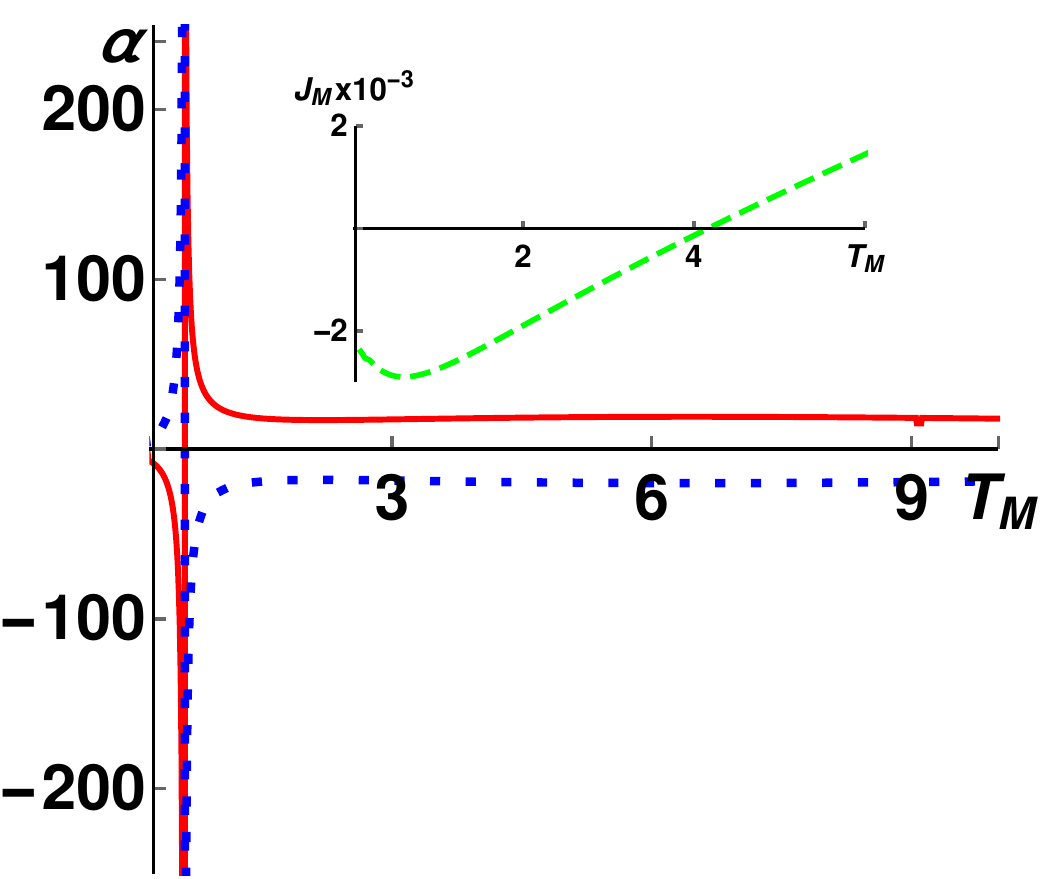}
\caption{\label{fig:3}
(Color online) The amplification factors defined in \Eq{\ref{alpha}}  as a function of the control parameter $T_M$.
The solid red line refers to $\alpha_L$ and the blue dotted one to $\alpha_R$.
In the inset, a zoom of $J_M$  including its local minimum responsible of
the divergence of the amplification factors.
The parameters have the same values of Fig.~\ref{fig:2}.}
\end{figure}

In order to better  describe
the emergence of the thermal transistor effect, the amplification factors $\alpha_s$
are plotted in Fig.~\ref{fig:3}, where it is possible to
see how the two lateral currents are amplified.
Notably, around $T_M \simeq 0.600$, the amplification factor
diverges, due to a local minimum of $J_M \simeq - 2.90 \times 10^{-3}$.
To this extent, we can circumscribe the  best working region of
this quantum thermal transistor to an interval
of the control quantity centered around this value.
However, a plateau is present where the amplification factors have values still significantly
high. In particular,  at $T_M \simeq 4.17$, where $J_M=0$,  $ \alpha_L \simeq 18.3$.
As a commentary, we notice that for this value of $T_M$, the reservoir $M$ does not need to inject  or absorb energy
to maintain the system in the steady state, or
in other words we can say that
the central reservoir is in a situation of dynamical thermal equilibrium
between the hot and the cold heat baths.

We also remark that it is easy to see that the right qubit is the one
showing $\chi_R < 0$. We have thus shown how a purely quantum
scenario can be used to implement a negative differential resistance device.
Other choices of the values of the parameters entering
either in  $H_S$ and in the dissipator of  \Eq{\ref{eq:master}} give
a behavior of the currents and of the amplifications analogous to the ones in Figs.~\ref{fig:2} and \ref{fig:3}.
This has been observed in the case of asymmetric configurations.

\subsection{Thermal transistor in a $z-$axis Ising model}
In order to better assess the value of the model  here proposed,
we compare it with the model previously studied in \cite{QTT}.
There, the thermal transistor effect had been observed in a range
of temperatures much narrower than in our case. We put ourself in the
same zone of parameters identified in Fig.~\ref{fig:2} and we
look if it is possible to extend the working region to larger zones
of temperatures as in the $x-$axis case.

\begin{figure}[t]
\includegraphics[height= 0.65 \columnwidth]{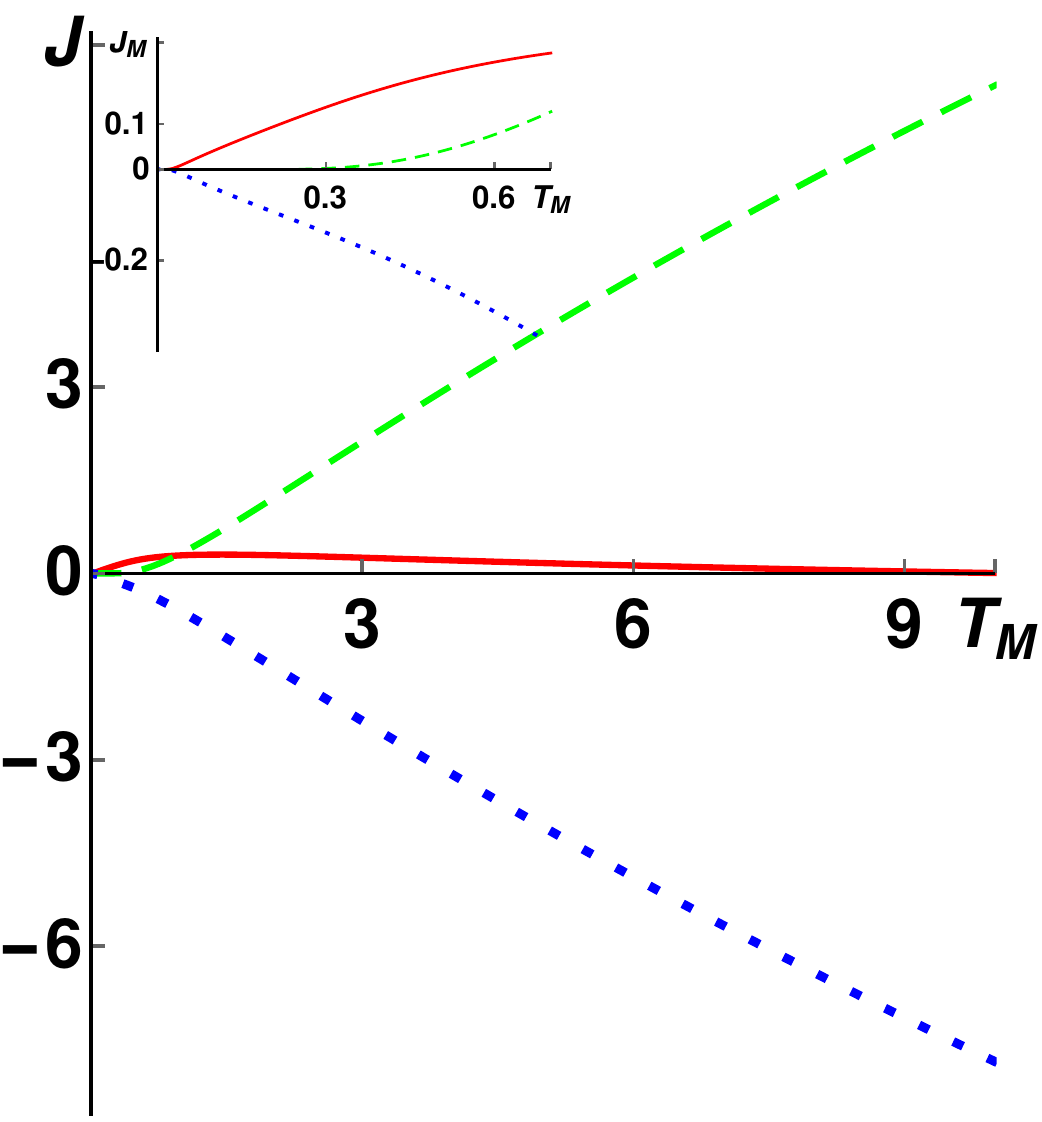}
\caption{\label{fig:4}
(Color online) The three thermal currents defined in \Eq{\ref{currJ}} exchanged by the
environments with the system as a function of the control parameter $T_M$,
when the system is described by a $z-$axis Ising model.
The solid red line refers to $J_L$, the dashed green line to $J_M$ and the blue dotted one to $J_R$.
The parameters have the same values of Fig.~\ref{fig:2}.
In the inset, a zoom including the working region of this setting.}
\end{figure}

In Fig.~\ref{fig:4}, we show that in the case of the $z-$axis Ising model
the thermal transistor effect works only in a narrow interval of the control temperature, namely $T_M \in [0, 0.2]$.
When moving to higher values of temperature, the effect completely vanishes,
no matter how large the thermal gradient between the hot and cold heat baths is.
In fact, in our analysis we have also increased the difference $T_R- T_L$
in comparison to what has been shown in \cite{QTT}  and observed that the current
amplification is always achieved in the same narrow working region.

Recently, the appearance of current amplification has been
discussed in a model based on a qubit-qutrit system. Also in this case,
the thermal transistor works in a narrow interval of the control temperature
in comparison to the $x$-type Ising model analyzed in Sec.~\ref{sec:IIIB}, \cite{QTT2}.

\subsection{Local behavior}

Here, we discuss the possibility to interpret the results of Fig.~\ref{fig:2}
on the basis of  considerations involving only the local behavior of the qubits composing the spin ring.
Firstly, we can have access to some local features
introducing the reduced density matrix of any qubit,
$\rho_p= \text{Tr}_{q,r \neq p}(\rho_{ss})$, where $\rho_{ss}$ is the global steady state.
We indicate with  $\rho_p^{1}$ and $\rho_p^{0}$, respectively,
the populations of the excited and ground state of the $p$-th qubit.
Each $\rho_p$ is a mixed state in the local energy basis and
such that we can define a local
temperature for each qubit $(\rho_p^{0}\ge \rho_p^{1}),$  with respect to their free Hamiltonianan, as:
 \begin{equation}
\label{Tloc}
T_p^\text{loc} =  \frac{\omega_p}{\ln ( \rho_p^{0}/ \rho_p^{1})}.
\end{equation}

\begin{figure}[t]
\includegraphics[width= 0.48 \columnwidth]{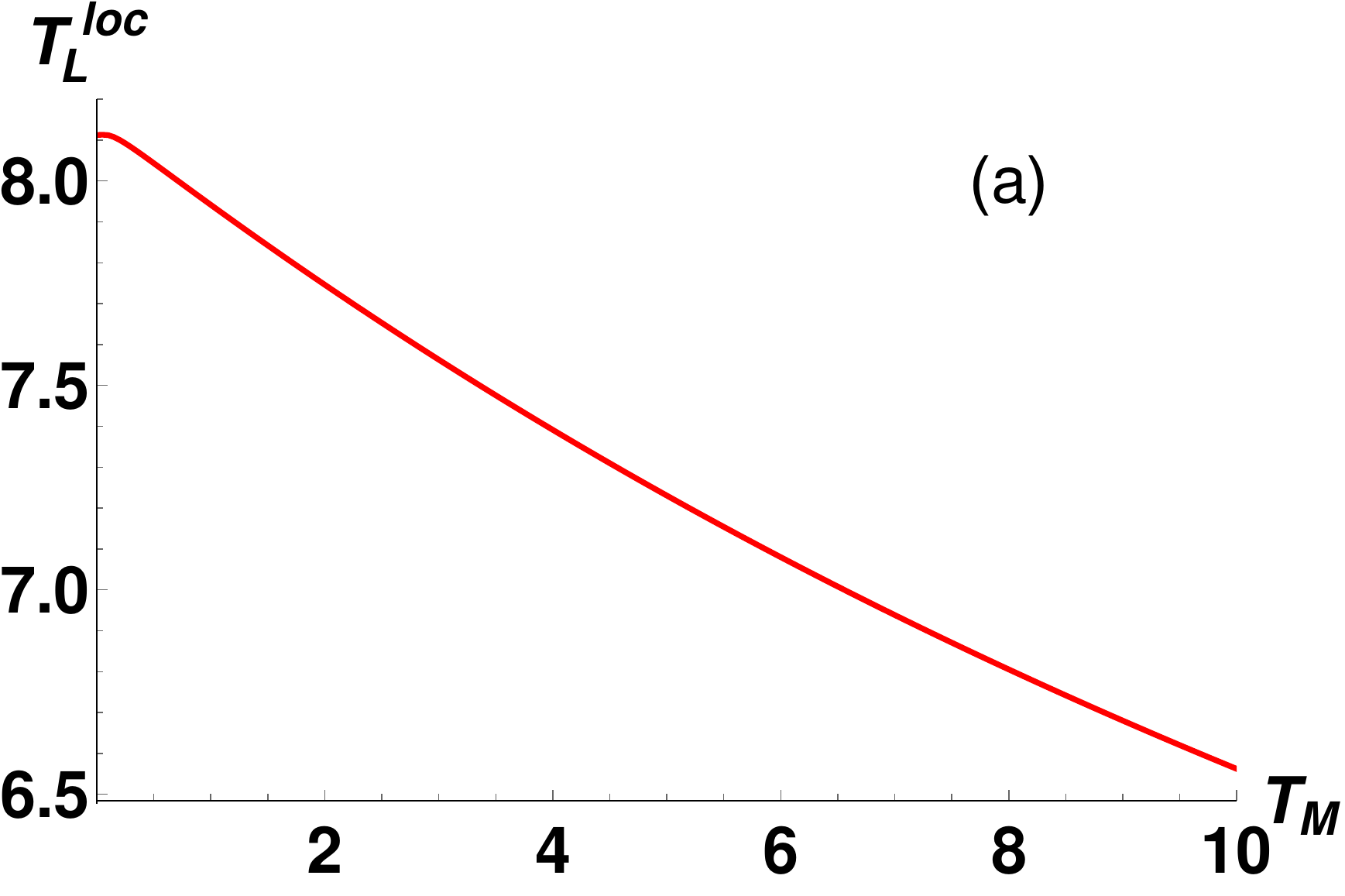}
\includegraphics[width= 0.48 \columnwidth]{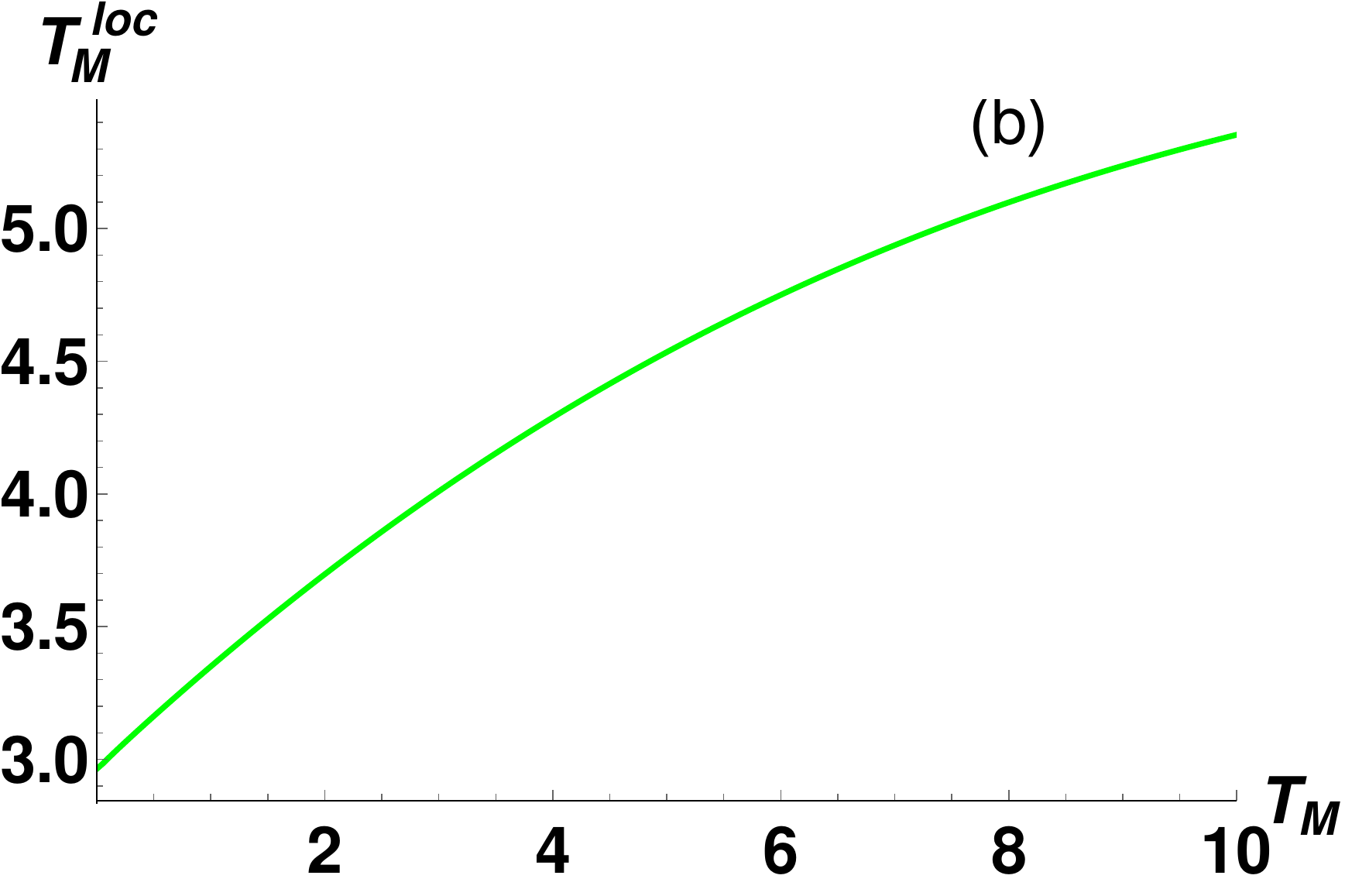}
\includegraphics[width= 0.48 \columnwidth]{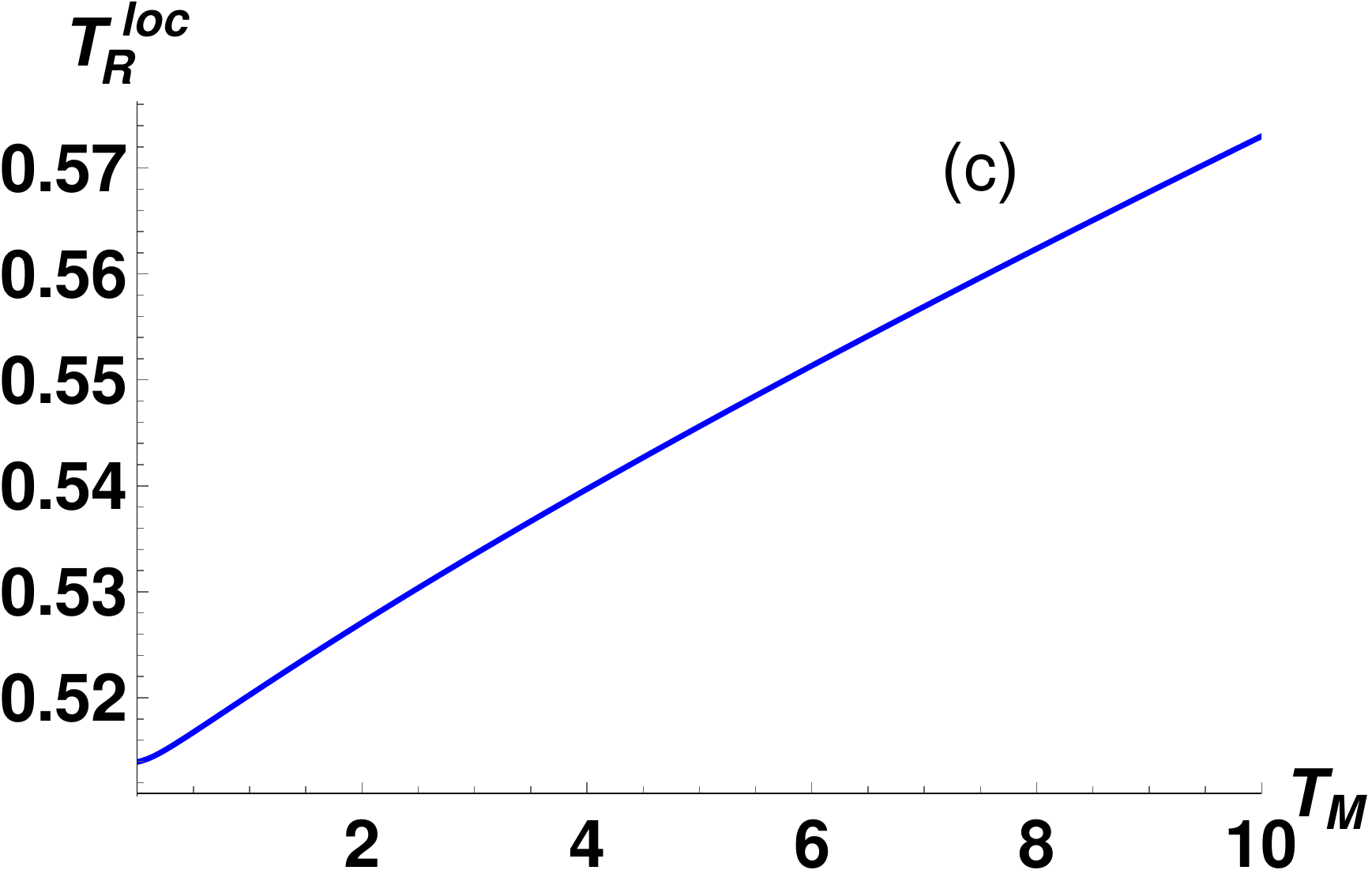}
\caption{\label{fig:5}
(Color online) Local temperatures  of the three qubits as defined
in \Eq{\ref{Tloc}}, as a function of the control
parameter $T_M$. Plots (a), (b) and (c) refer, respectively, to
the local temperature of the left, middle and right qubit.
The parameters have the same values of Fig.~\ref{fig:2}.}
\end{figure}

 We plot in Fig.~\ref{fig:5} the local temperatures of the qubits as a function of the control temperature $T_M$.
It is clear that by increasing  $T_M$, the temperature gradient
between the $L$ and $M$ qubits decreases, in contrast to what happens
between the pair of qubits $M$ and $R$.
In addition, the external qubits are never in thermal equilibrium
with their respective environments, only the middle one
reaches the equilibrium with the control thermostat $M$ at $T_M \simeq 4.17$.
For this value  it holds $J_M = 0$ and $J_L = - J_R \simeq 0.179 $.
This means that all the heat injected by the hot reservoir into the system is transferred
directly to the cold bath without any participation of the control bath.

Once analyzed the behavior of the local temperatures, we examine
if we can qualitatively reproduce the behavior of the heat currents with
the following local model. The three qubits are in contact with their local
thermostats exactly as in our model, but the qubits do not interact between themselves.
Instead, each qubit is also strongly coupled to another local thermal environment
whose temperature is equal to (for any value of $T_M$)
the temperature $T_p^\text{loc}(T_M)$ depicted in Fig.~\ref{fig:5}.
The steady state of each qubit can be made close at will to the thermal
state of temperature $T_p^\text{loc}(T_M)$ by suitably tuning the coupling
between qubits and their own two local reservoirs. In this local model,
the reduced state of each qubit is thus equal to the one obtained in our model,
while the way each qubit dissipates is strongly different. In the local model the
dissipation is governed by local Lindblad operators while in our model
by collective ones, computed using the dressed states of $H_S$.

\begin{figure}[t]
\includegraphics[width= 0.48 \columnwidth]{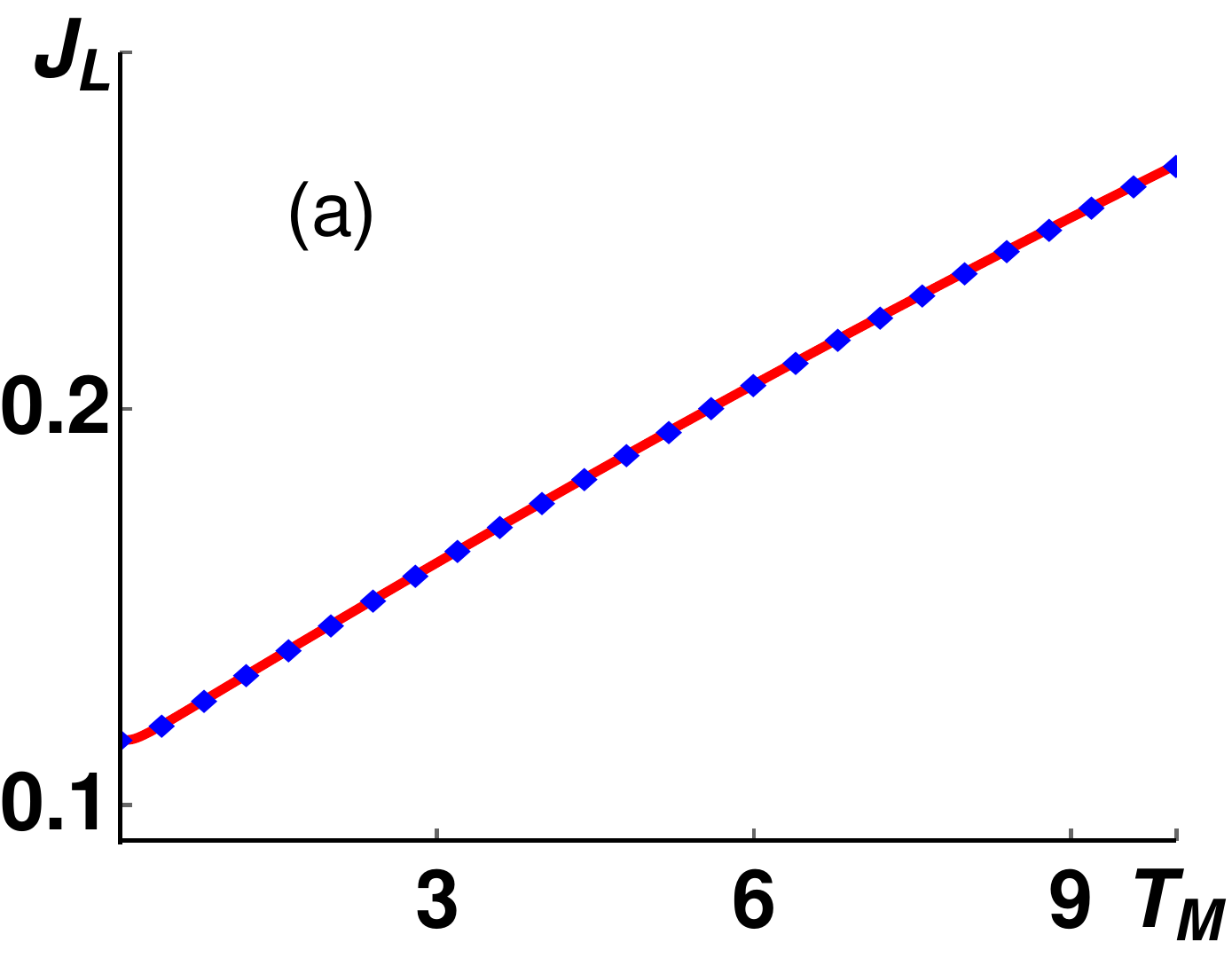}
\includegraphics[width= 0.48 \columnwidth]{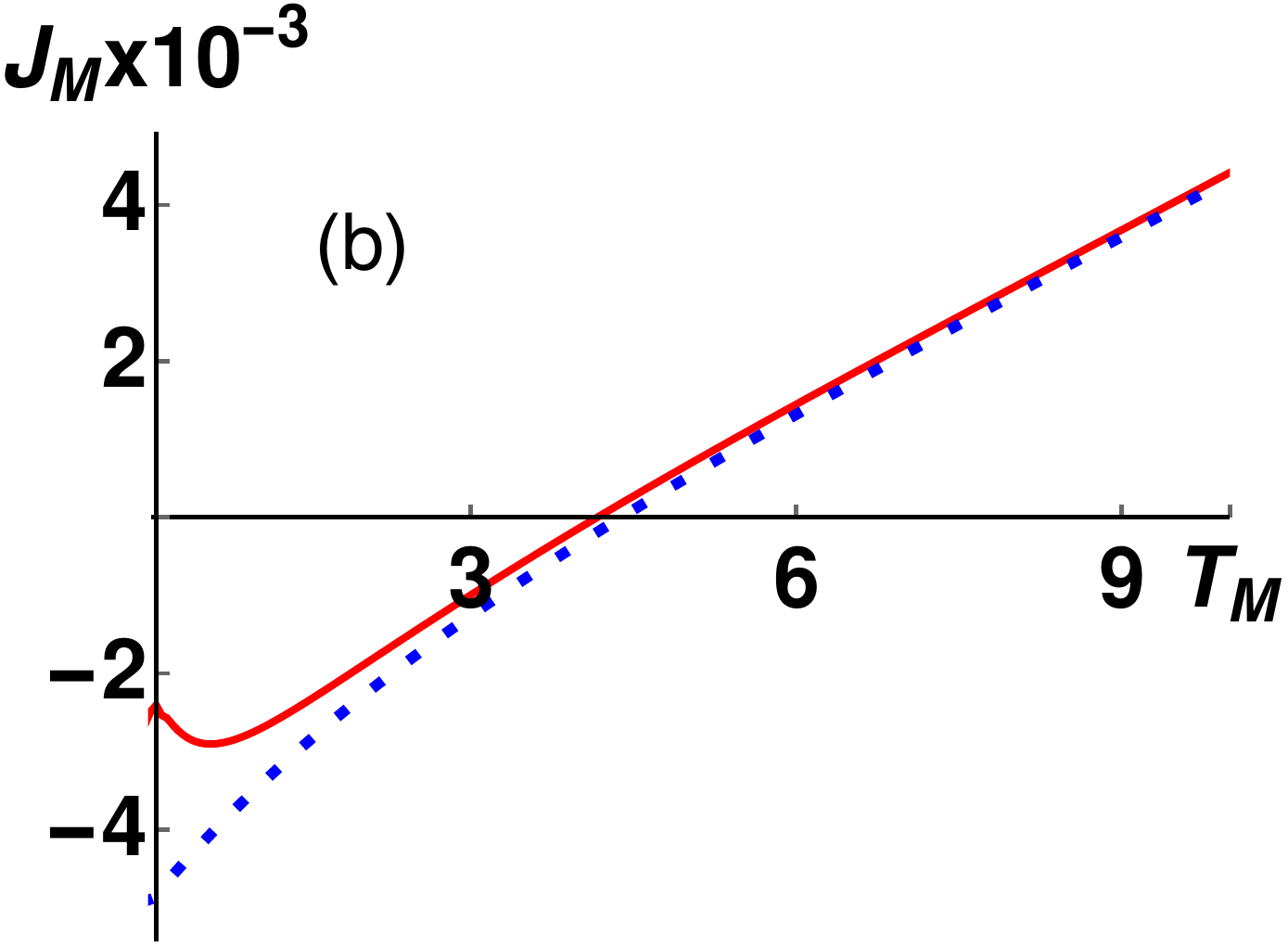}
\includegraphics[width= 0.48 \columnwidth]{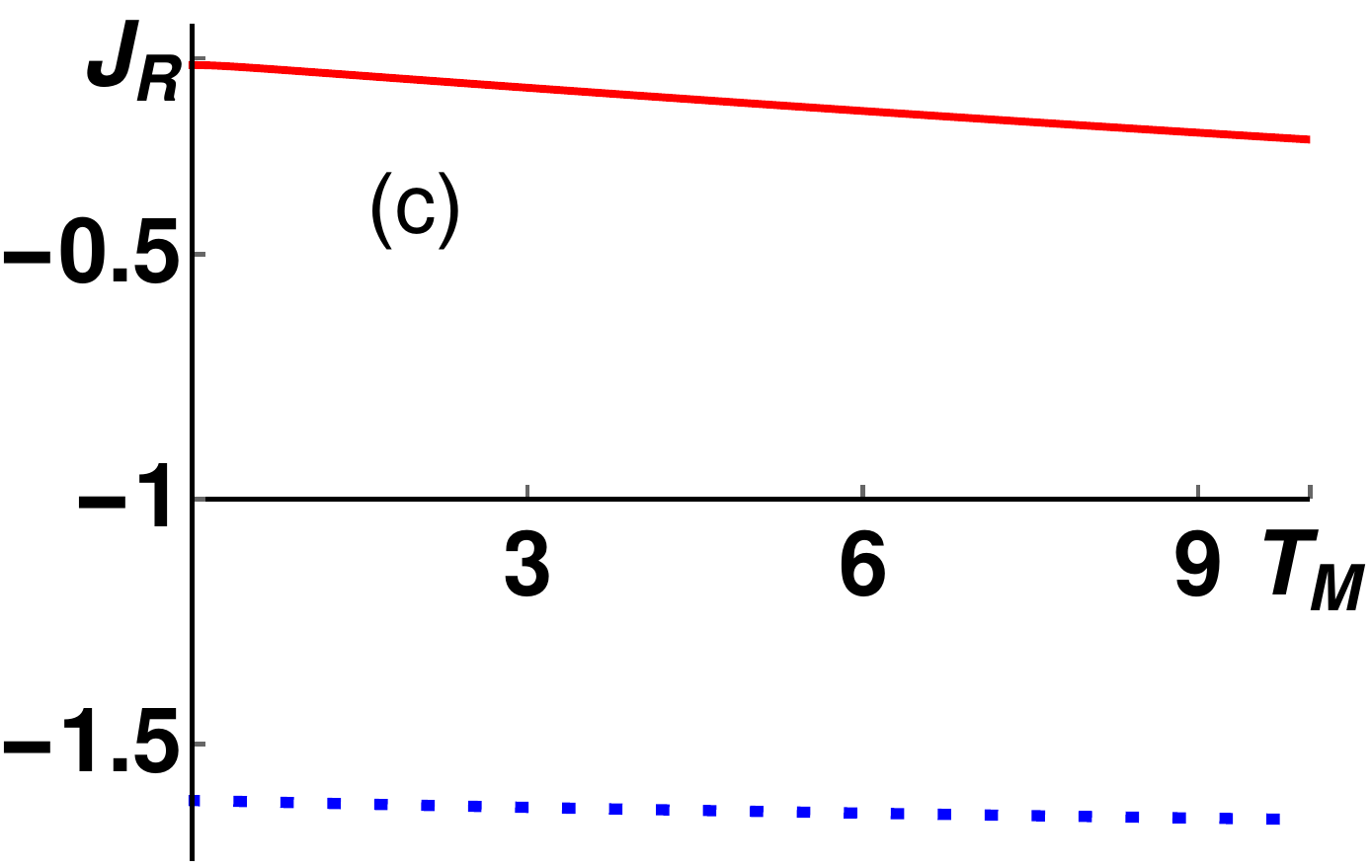}
\caption{\label{fig:6}
(Color online)  Comparison between the exact currents
computed for the $x$-axis Ising model (red lines), the same ones of Fig.~\ref{fig:2}, and the
currents obtained with the local model (dotted  blue lines).}
\end{figure}

In Fig.~\ref{fig:6}  we compare the local currents with the exact ones of Fig.~\ref{fig:2}.
We notice that in some cases, local currents well approximate the exact ones, like $J_L$,
while in other they do not, like $J_R$. This is a clear indication of the relevant role
played by the collective behavior of the spin ring when it exchanges heat
with the local thermostats.
The local temperatures induced by the collective dynamics
are thus, in general, not enough by themselves to catch all the physics of
this collective exchange and of the occurrence of the thermal transistor effect. 
In particular, we stress the absence of a local minimum for $J_M$, around which we observed the best performance of the device.

\section{Robustness against perturbations}
\label{Robust}

The results reported in the previous sections show how
to build a quantum thermal transistor operating on
a wide range of temperatures and starting from
a microscopic derivation, with a potential
implementation over some feasible physical settings.
However, in any real physical implementation of an Ising Hamiltonian,
spurious and undesired couplings pointing along different
directions of the  desired axis for the interaction could occur. This could entail
a lost of the thermal transistor effect. Obviously, in any controlled  setup, those
interactions can be considered as small fluctuations and treated
by means of perturbative corrections to the system Hamiltonian $H_S$.
Nevertheless, in order to pursue our microscopical description
of the effect and be precise, we use the full model
for the three qubits as in \Eq{\ref{eq:hsys}}, introducing
time by time more complex interactions and studying
to which extent they alter the amplification performances.

In order to substantiate the process illustrated so far,
we consider a physical implementation where the  real axis of the interaction term
of the model can differ from the expected one
(indicated with $k$ in the following) \cite{ExpIsing1},
and we compute the amplification factors, $\alpha_s(\lambda^{i},\lambda^{j})$,
that are now functions of the couplings, $\lambda^{i}$ and $\lambda^{j}$, along the other
two possible polarization directions $i$ and $j$,  and they
may assume a different value from the unperturbed one.
Only the coupling between the qubits $L$ and $M$  and
between the qubits $R$ and $M$ are perturbed,
while the  coupling between the qubits $L$ and $R$ is maintained equal to zero as before.

To obtain a statistical detailed analysis we assume that the
values of the perturbations are normally distributed and we evaluate the
averaged amplification factors
\begin{equation}
\label{alphaAv}
\overline{\alpha}^{i,j}_s= \int d\lambda^{i} d\lambda^{j} \alpha_s(\lambda^{i},\lambda^{j})
\boldsymbol{\mathcal{N}}(\overline{\lambda}^{i,j},\Sigma^{i,j}),
\end{equation}
where $s = L, R$, and $\boldsymbol{\mathcal{N}}(\overline{\lambda}^{i,j},\Sigma^{i,j})$
is the bivariate Gaussian distribution of the perturbations.
It is centered around their mean values
$\overline{\lambda}^{i,j} = (\overline{\lambda}^{i},\overline{\lambda}^{j} )$,
with covariance matrix given by
$ \Sigma^{i,j} = \text{diag}(\Sigma^{i},\Sigma^{j})$.
In particular, we choose that the spurious interactions can
deviate of  around $10\%$  from their mean value,
i.e. $\Sigma^{i}= 0.1 \overline{\lambda}^{i}.$

\begin{figure}[t]
\includegraphics[width= 0.75 \columnwidth]{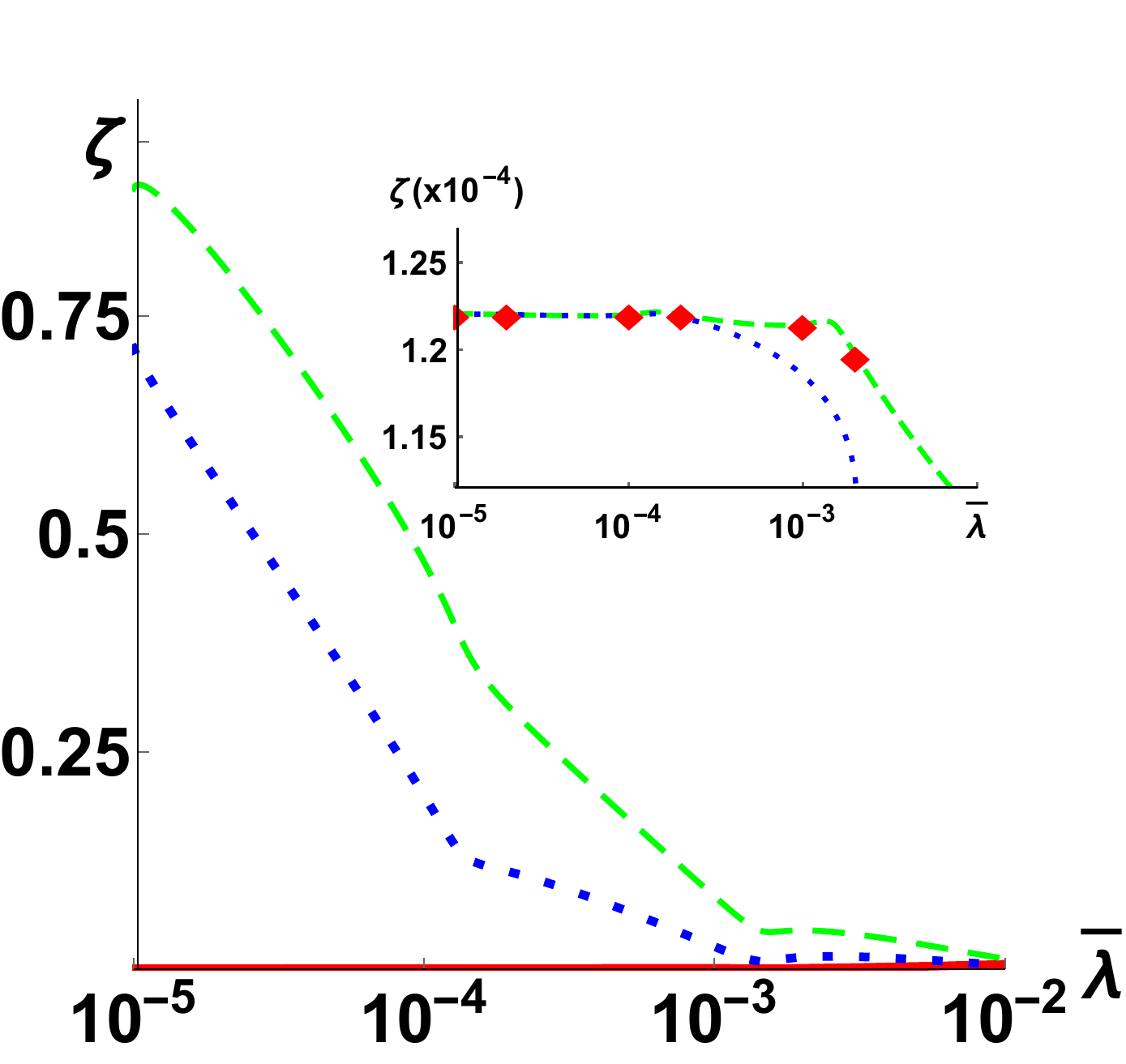}
\caption{\label{fig:7}
(Color online) The ratio $\zeta$ defined in \Eq{\ref{ratio}} for
the $x-$type Ising model, as a function of the mean value of
the spurious couplings  $\overline{\lambda}^{i,j}$.
The dashed green line refers to  $\zeta^{z}$, the blue dotted one
to $\zeta^{y}$ and the red solid line takes into account
perturbations with both  $\lambda^{y} $ and  $\lambda^{z} \neq 0$ but with the same mean value, namely $\zeta^{y,z}$.
In the inset, the ratio $\zeta$ for the $z-$type Ising model.
Here, the dashed green line refers to $\zeta^{x}$, coinciding with $\zeta^{y}$  (red diamonds).
The blue dotted one refers to $\zeta^{x, y }$ and is\ obtained
imposing $\lambda^{y}$ and $ \lambda^{z} \neq 0$ with the same mean value.
The parameters of the unperturbed configuration are the same as in  Fig.~\ref{fig:2}.}
\end{figure}

As a tool to investigate the effect of small perturbations, we define the ratio
\begin{equation}
\label{ratio}
\zeta^{i,j} = \frac{\overline{\alpha}_{L}^{i,j}}{\alpha_{L}^{\text{unp}}},
\end{equation}
where $\alpha^{\text{unp}}_{L}$ represents the value of the left amplification factor
obtained in the absence of perturbation.  If only  one additional coupling of the type
$\sigma_p^i \sigma_q^i $ perturbs the main interaction
term $\sigma_p^k \sigma^k_q $, the above ratio is indicated with $\zeta^{i}$.
For the sake of clarity, we point out that another quantity analogue to $ \zeta^{i,j}$ can be
in principle defined using the amplification factor of the right current $\alpha_{R}$.
Nevertheless, its value depends  on $\alpha_{L}$ in a simple way, such that for any  $T_M$
this choice would lead to analogues dampings of $ \zeta^{i,j}$ as function of the
strength of the perturbation, $\overline{\lambda}^{i,j}$.

In Fig.~\ref{fig:7} we show the quantity $\zeta$
as a function of the average strength of 3 types of typical
perturbations that can occur in both settings discussed in Section $\ref{QTT}$.
Since $\alpha^{\text{unp}}_{L}$ can diverge for a  certain value of the control temperature $T_M$,
we fix it equal to $T_M = 0.650$, close enough the one where the divergence appears 
in order to have $ \alpha^{\text{unp}}_{L}\gg1$ but where it is still finite, being $\alpha_L^{\text{unp}} \simeq 1.42 \times 10^3$.
We stress here that  the value of $T_M$ in the presence of perturbations is kept the same of the unperturbed case. 
The  quantity $\zeta$ is close to unity when the
system still operates as a quantum transistor close to its unperturbed value
$\alpha^{\text{unp}}_L$ but it vanishes as soon as
the perpendicular couplings become more intense.

As it is apparent from the plot, the $x-$axis and $z-$axis Ising  models
show a completely opposite behavior in  the presence of perturbations.
Indeed, despite the fact that both the models have been perturbed around a quite
optimal  configuration, the $x-$axis model still shows the thermal transistor effect
for extra couplings small enough.
In contrast, the possibility to achieve a current amplification with the
$z-$axis Ising model is completely lost in  the presence of  infinitesimal couplings
among the spin in any direction perpendicular to  the wanted interaction axis.
Our numerical evidences only allow us to see that $\zeta$ is extremely low
for values of the extra couplings larger than ten millionths
of the one along the $z-$axis, but we cannot surmise how
the unperturbed value is restored for smaller values of the spurious coupling $\overline{\lambda}^{i,j}$.

We also provide a qualitative mathematical explanation of this phenomenon
which resides on the structure of the couplings between the qubits.
In fact, as pointed out when discussing the two possible configurations,
a coupling commuting with the bare Hamiltonian
affects only the eigenfrequencies of the systems and it
leaves unchanged the eigenstates entering in the
definition of the operators $A(\omega)$, that are essential
to derive the dissipative dynamics of \Eq{\ref{eq:master}}.
Adding a perpendicular coupling, not commuting
with the bare part of $H_S$ completely alters their structure and it
imposes to consider the dressed-state basis to write the $A(\omega)$,
that typically differs from the bare-state basis one.

\section{Conclusions}
\label{Concl}

In this paper, we have addressed the design and the
possible physical implementation of a quantum thermal transistor.
This novel device acts on thermal currents as a bipolar electronic transistor
does with electric currents, providing an amplification of the
collector's and emitter's currents by a proper tuning of the gate
potential, while the corresponding gate current is
order of magnitude lower than the other two and practically constant.

Our model proposes a system of three qubits strongly interacting
among them and each of them in contact with a different thermal reservoir. This setting imposes non-equilibrium
dissipative dynamics that entail the presence of
non-vanishing steady state currents flowing into the system via bath induced transitions.
We have focused our attentions on a particular system that
can lead to a feasible realization on several experimental platforms,
such as molecular nanomagnets and few-body spin chains,
and we have referred to this model as the spin-ring model.

In particular, we have devoted our efforts to describe how a
quantum thermal transistor works in a scenario where a more
complex spin-spin coupling is present, in comparison to a previous model
recently presented in \cite{QTT}.
Throughout the paper, we have referred to these two models as the $x$-type Ising
model and the $z$-type Ising model. The former is characterized
by an interaction Hamiltonian not commuting with the bare one, or in other
words the magnetic  field acting on the three spins is perpendicular
to the direction along which the spins interact; the latter, instead,
is a model where magnetic fields and interaction are longitudinal,
so that the two contributions of the system Hamiltonian commute.

Choosing as control parameter the temperature of the heat bath directly
coupled to the middle qubit, we have shown that the $x$-type spin ring operates
as a thermal transistor over a wider range of temperatures in comparison to the $z$-type one.
The significant amplification of the lateral currents has
been justified in the context of devices characterized by
negative differential thermal resistance \cite{LLC06, graphene, silicene}
and by means of the collective dynamics. In fact, considering a naive local
model for the system, that partially takes into account collective phenomena,
cannot explain the whole behavior exhibited by our model.
Finally, we have discussed the robustness of both the possible
implementations against spurious and uncontrollable couplings stemming
from the magnetic interaction between the different pairs of the effective
spins of the magnetic ring. The results of exact  numerical computations suggest that,
for reasonably small  perturbations, the $x$-type molecule can be steered acting on
the control temperature to behave still as a thermal transistor around its unperturbed configuration.
In contrast, our computations show that the $z$-type molecule
is really fragile in presence of those kinds of perturbations.

As final remarks, we would like to point out that even
if we have performed a microscopic derivation of master
equation describing the dynamics of the system, overriding any
phenomenological simplification, a deeper comprehension of the very
quantum effects conducting to such current amplification is still lacking.
It  could be matter of future investigation the role of system criticality
and of the dissipation mechanism and how they contribute and compete
in order to state a general scheme to build a quantum thermal transistor.

\begin{acknowledgments}
A.M. acknowledges the ``R\'{e}gion
Bourgogne-Franche-Comt\'{e}'' for financial support
throughout the mobility grant N. 2017Y-06400. 
B.B. acknowledges the support by the French ``Investissements d'Avenir'' program, project ISITE-BFC (contract ANR-15-IDEX-03).
A.M. thanks Heinz-Peter Breuer for useful comments and Elena Servida for discussions
about semiconductor electronics. B.B. thanks Benedetto Militello and David Viennot  for helpful comments.
\end{acknowledgments}

\appendix*

\section{Spectrum of the system Hamiltonian  in the $x$-type Ising model.}
\label{ApA}

In this appendix we give a sketch of the diagonalization
of the system Hamiltonian introduced in  \Eq{\ref{eq:hsys}} and used to
obtain the thermal transistor effect discussed throughout the paper.
In particular, we focus on the case of the $x$-type Ising model with  
the left and right qubits not interacting, i.e. $\eta_{LR}=0$.
The matrix of a system composed by three qubits with different
splitting in a magnetic field with a transverse Ising interaction reads:

\begin{widetext}
\begin{equation}
\label{eq:matrix}
\frac12 \left(
\begin{array}{cccccccc}
 \Omega_{LM}+\omega_R & 0 & 0 & \Lambda_{MR}^{x} & 0 & 0 & \Lambda_{LM} ^{x} & 0 \\
 0 &  \Omega_{LM}-\omega_R & \Lambda_{MR} ^{x} & 0 & 0 & 0 & 0 & \Lambda_{LM} ^{x} \\
 0 & \Lambda_{MR} ^{x} &  \Omega_{LR}-\omega_M & 0 & \Lambda_{LM} ^{x} & 0 & 0 & 0 \\
 \Lambda_{MR} ^{x} & 0 & 0 & \omega_L-\Omega_{MR} & 0 & \Lambda_{LM} ^{x} & 0 & 0 \\
 0 & 0 & \Lambda_{LM} ^{x} & 0 & -\omega_L+ \Omega_{MR} & 0 & 0 & \Lambda_{MR} ^{x} \\
 0 & 0 & 0 & \Lambda_{LM} ^{x} & 0 & \omega_M- \Omega_{LR} & \Lambda_{MR} ^{x} & 0 \\
 \Lambda_{LM} ^{x} & 0 & 0 & 0 & 0 & \Lambda_{MR} ^{x} & - \Omega_{LM}+\omega_R & 0 \\
 0 & \Lambda_{LM} ^{x} & 0 & 0 & \Lambda_{MR} ^{x} & 0 & 0 & - \Omega_{LM}-\omega_R \\
\end{array}
\right),
\end{equation}
\end{widetext}
where we have assumed $\eta_{qp}=\eta_{pq}$ and for shorthand we have introduced
$ \Omega_{pq}= \omega_p + \omega_q$ and $\Lambda_{pq}^{x}= \eta_{pq} \lambda^{x}$ with $ p, q  = L, M, R $.

The secular equation of \Eq{\ref{eq:matrix}} can be reduced to a quartic
equation, leading to four pairs of  energy eigenvalues equal in modulus.
We can, therefore, decompose $H_S$ in the base where it is diagonal,
$H_s = \sum_{i=1}^8  \epsilon_i \ketbra{\epsilon_i}{\epsilon_i}$, where we have ordered the
energy eigenvalues using the following notation
$ \epsilon_8 \leq \epsilon_7 \leq ... \leq \epsilon_1  $ and $ \epsilon_{9-i} = - \epsilon_i, \, \forall i $.

Introducing the computational basis, defined as
the tensor product $\ket{l_{L}l_{M}l_{R}}$, of all the dispositions of the eigenvectors of
the three $\sigma_p^z$, i.e.  $\sigma_p^z \ket{l_p}=(-1)^{(l_p+1)}~\ket{l_p}$,
with $l_p= 0,1 $ and $ p= L, M, R $, the eigenvectors of $H_S$,  found for the main case discussed in the paper (Sec.~\ref{sec:IIIB}), are:
\begin{equation}
\begin{split}
\label{eigenvec}
\ket{\epsilon_1} &= -a_1 \ket{111} -a_2 \ket{100} -a_3 \ket{010} -a_4 \ket{001} \\
\ket{\epsilon_8} &= +a_1 \ket{000} -a_2 \ket{011} +a_3 \ket{101} -a_4 \ket{110} \\
\ket{\epsilon_2} &= + b_1 \ket{000} +b_2 \ket{011} +b_3 \ket{101} +b_4 \ket{110} \\
\ket{\epsilon_7} &= + b_1 \ket{111} -b_2 \ket{100} +b_3 \ket{010} -b_4 \ket{001} \\
\ket{\epsilon_3} &= +c_1 \ket{000} -c_2 \ket{011} -c_3 \ket{101} +c_4 \ket{110} \\
\ket{\epsilon_6} &= -c_1 \ket{111} -c_2 \ket{100} +c_3 \ket{010} +c_4 \ket{001} \\
\ket{\epsilon_4} &= +d_1 \ket{111} -d_2 \ket{100} -d_3 \ket{010} +d_4 \ket{001} \\
\ket{\epsilon_5} &= -d_1 \ket{000} -d_2 \ket{011} +d_3 \ket{101} +d_4 \ket{110}
\end{split},
\end{equation}
where the 16 real coefficients $a_i, b_i, c_i, d_i$ with $i= 1,2,3,4 $
in the above expressions  are functions of the parameters
entering in the Hamiltonian.
In all the numerical trials, we have found the same above structure for the eigenvectors, all the coefficients being real.

It is worth noting that the above states have well-defined parity and that the states with opposite energy have
different parity. In fact, it is easy to check it by measuring
the parity operator  $P= \sigma_L^z \otimes \sigma_M^z \otimes \sigma_R^z $ on any term
appearing in $\ket{\epsilon_i}$.

Finally, we report in Table~\ref{t:transition} the transitions mediated by any thermal reservoir.
It is worth noting that we write here only the transitions that can be
interpreted as the emission of a photon in the usual context of quantum optics,
since they involve states with a positive energy difference  between the initial and the final value.

\begin{table}[h]
\caption{\label{t:transition}
The possible transitions of positive frequency
mediated by any of the three thermal baths  by means of the interaction operator $\sigma_p^x$ for the case of Sec.~\ref{sec:IIIB}.
The first column gives the initial states, starting from the one
with highest energy, whilst the other columns specify 
the qubit operator coupled to the corresponding thermal
reservoir as prescribed by \Eq{\ref{eq:hamiltonians}}, and the final states connected to the initial ones.}
 \vspace{2mm}
\begin{ruledtabular}
\begin{tabular}{c | c |  c | c }
\quad &  $\sigma^x_L $ & $\sigma^x_M$ & $\sigma^x_R$ \\[1mm] \hline
$\ket{\epsilon_1}$ & $\ket{\epsilon_2}$, $\ket{\epsilon_3}$, $\ket{\epsilon_5}$ & $\ket{\epsilon_2}$, $\ket{\epsilon_3}$, $\ket{\epsilon_5}, \ket{\epsilon_8}$ & $\ket{\epsilon_2}$, $\ket{\epsilon_3}$, $\ket{\epsilon_5}$  \\[.8mm] \hline
$\ket{\epsilon_2}$& $\ket{\epsilon_4}$, $\ket{\epsilon_6}$& $\ket{\epsilon_4}$, $\ket{\epsilon_6}, \ket{\epsilon_7}$ & $\ket{\epsilon_4}$, $\ket{\epsilon_6}$ \\[.8mm]  \hline
$\ket{\epsilon_3}$&  $\ket{\epsilon_4}$, $\ket{\epsilon_7}$& $\ket{\epsilon_4}$, $\ket{\epsilon_6}, \ket{\epsilon_7}$&  $\ket{\epsilon_4}$, $\ket{\epsilon_7}$ \\[.8mm] \hline
$\ket{\epsilon_4}$ &  $\ket{\epsilon_8}$& $\ket{\epsilon_5}, \ket{\epsilon_8}$ & $\ket{\epsilon_8}$ \\[.8mm] \hline
$\ket{\epsilon_5}$ &  $\ket{\epsilon_6}$, $\ket{\epsilon_7}$& $\ket{\epsilon_6}, \ket{\epsilon_7}$ & $\ket{\epsilon_6}, \ket{\epsilon_7}$ \\[.8mm]\hline
$\ket{\epsilon_6}$ &  $\ket{\epsilon_8}$& $\ket{\epsilon_8}$ & $\ket{\epsilon_8}$ \\[.8mm]\hline
$\ket{\epsilon_7}$ &  $\ket{\epsilon_8}$& $\ket{\epsilon_8}$& $\ket{\epsilon_8}$ \\[.8mm]
\end{tabular}
\end{ruledtabular}
\end{table}

\end{document}